\documentclass[preprint]{aastex63}
\usepackage{color}
\usepackage{bm}
\usepackage{cases}
\usepackage{multirow}
\usepackage{lineno}

\received{}
\revised{}
\accepted{}
\submitjournal{APJ}
\shorttitle{}
\shortauthors{}
\begin{document}

    \title{Redshift-evolutionary X-Ray and UV luminosity relation of quasars from Gaussian copula}

    \author{Bao Wang}
      \affiliation{Department of Physics and Synergistic Innovation Center for Quantum Effects and Applications, Hunan Normal University, Changsha, Hunan 410081, China}

    \author{Yang Liu}
      \affiliation{Department of Physics and Synergistic Innovation Center for Quantum Effects and Applications, Hunan Normal University, Changsha, Hunan 410081, China}

   \author{Zunli Yuan}
  \affiliation{Department of Physics and Synergistic Innovation Center for Quantum Effects and Applications, Hunan Normal University, Changsha, Hunan 410081, China}

  \author{Nan Liang}
  \affiliation{Key Laboratory of Information and Computing Science Guizhou Province, Guizhou Normal University, Guiyang, Guizhou 550025, People's Republic of China}

    \author{Hongwei  Yu}
    \affiliation{Department of Physics and Synergistic Innovation Center for Quantum Effects and Applications, Hunan Normal University, Changsha, Hunan 410081, China}

    \author{Puxun Wu}
    \affiliation{Department of Physics and Synergistic Innovation Center for Quantum Effects and Applications, Hunan Normal University, Changsha, Hunan 410081, China}

    \email{hwyu@hunnu.edu.cn}
   \email{liangn@bnu.edu.cn}
   \email{pxwu@hunnu.edu.cn}

\begin{abstract}

We construct a three-dimensional and redshift-evolutionary   X-ray and ultraviolet ($L_X-L_{UV}$) luminosity relation for   quasars from the powerful statistic tool called copula,  and find that the constructed $L_X-L_{UV}$ relation from copula is more viable than the standard one and the observations favor the redshift-evolutionary relation more than $3\sigma$. The Akaike and Bayes information criterions indicate that the quasar data support strongly the three-dimensional $L_X-L_{UV}$ relation. Our results show that  the quasars can be regarded as a reliable  indicator of the cosmic distance if the  $L_X-L_{UV}$ relation from copula is used to calibrate quasar data.

~\\

\end{abstract}

\section{Introduction}

Quasars (quasi-stellar objects, QSOs) are the extremely luminous persistent sources in  the universe. They are found  in the centers of some galaxies and powered by gas spiraling at high velocity into an extremely large black hole.   The most powerful quasars have luminosities thousands of times greater than that of the Milky Way, and thus quasars can be visible at the very remote distance. The redshift of a quasar can reach $z>7$~\citep{Mortlock2011luminous, Ba2018million, Wang2021Luminous}. Today, more than half a billion  quasars have been identified by observations~\citep{Lyke2020}.   If quasars can be regarded as  standard candles for probing the cosmic evolution, they can  cover effectively the redshift desert of the cosmic observational data. Thus, quasars can help us to understand the properties of dark energy and the possible origin of the present  Hubble constant ($H_0$) tension.

 Properties of quasars have been proposed as probes to explore the cosmic expansion.   These include quasar angular size measurements \citep{Paragi1999VLBI, Chen2003Cosmological, Cao2017Ultra, Ryan2019Baryon, Cao2020Cosmological, Cao2021Cosmological}, the anticorrelation between UV emission lines and luminosity \citep{Baldwin1977Luminosity, Osmer1999Review}, the luminosity-mass relation in super-Eddington accreting quasars \citep{Wang2014Supermassive}, the relation between luminosity and X-ray variability \citep{La2014New} and the radius-luminosity relationship \citep{Watson2011New, Melia2014high, Kilerci2015Scatter}.

The nonlinear relation between the  X-ray luminosity and ultraviolet (UV) luminosity ($L_X-L_{UV}$) in quasars has been constructed to derive the distance of quasars and to build a quasar Hubble diagram  up to $z\sim 7$~\citep{Risaliti2015Hubble, Risaliti2019Cosmological, Lusso2016, Lusso2017,Lusso2020Quasars}.  The $L_X-L_{UV}$ relation has been used widely  in quasar cosmology \citep{Lusso2019,Wei2020,Khadka2020Quasar, Khadka2020Using, Khadka2021Determining, Lian2021Probing, Li2021Hubble, Bargiacchi2021Quasar,Hu2022}.  \cite{Risaliti2019Cosmological} found that there is a larger than 4$\sigma$ tension between the  1598  quasar data sets and the standard spatially-flat $\Lambda$CDM model with $\Omega_\mathrm{m0} = 0.3$, where $\Omega_\mathrm{m0}$ is the present dimensionless matter density parameter. This may be a sign that  quasars cannot be treated as the standard candle if the $L_X-L_{UV}$ relation is  used \citep{Khadka2020Using} and there have already been some discussions and controversies about this \citep{Yang2020,Banerjee2021,Dainotti2022, Singal2022, Petrosian2022, Li2022}. Furthermore, some data show evidence of redshift evolution of the X-ray and UV correlation \citep{Khadka2022quasar}, although recently \cite{Sacchi2022Quasars} found that this correlation still holds at $z>2.5$ by performing a one-by-one analysis of a sample of 130 quasars  with high-quality  X-ray and UV spectroscopic observations.  Therefore, it remains  interesting to research the possible redshift evolution of   the $L_X-L_{UV}$ relation.

Copula is a powerful tool developed in modern statistics to describe the correlation between multivariate random variables~\citep{nelsen2007introduction}),  and it has been used widely in various areas such as finance and hydrology. In recent years, copula has gradually been recognized by the astronomical community as an important tool to analyze data. 
For example, copula have been used to determine the luminosity function of the radio cores in the active galactic nucleus \citep{Yuan2018Determining}, construct a period-mass function for extrasolar planets \citep{Jiang2009Construction}, and build  the correlation
between the gamma ray burst (GRB) peak energy and the associated supernova peak brightness \citep{Koen2009Confidence} and a new
approximation for the low multipole likelihood of the cosmic microwave background (CMB) temperature \citep{Benabed2009TEASING}.
Furthermore, copula has been introduced to research the convergence power spectrum from the weak lensing surveys \citep{Sato2010, Sato2011}, the gravitational-wave astronomy \citep{Adamcewicz2022}, the galaxy luminosity functions \citep{Takeuchi2010, Takeuchi2013, Takeuchi2020} and the large-scale structure fields of matter density \citep{Scherrer2010, Qin2020}.
Recently, copula is used successfully in gamma ray burst (GRB) cosmology and gives an improved  Amati correlation of GRBs~\citep{Liu2022a}. Utilizing this improved Amati correlation to calibrate  GRBs, the GRB data can give  cosmological results, which are consistent with what were obtained from other popular data~\citep{Liu2022b}.

Thus, we expect that copula may also play an important role in constructing the relations of quasars.
In this work, we aim to construct, by using the powerful statistical tool copula, a three-dimensional  $L_X-L_{UV}$ relation, which contains  a redshift-dependent term. From the three-dimensional Gaussian copula, we obtain a three-dimensional and redshift-evolutionary $L_X-L_{UV}$ relation.  Comparing it with the standard $L_X-L_{UV}$ relation by using the latest X-ray and UV flux measurements data \citep{Lusso2020Quasars} that  contain 2421 data sets of measurements,  we find that the addition of the redshift-dependent term improves the viability  of the $L_X-L_{UV}$ correlation  significantly.

The rest of the Paper is organized as follows. In Section 2, we introduce the Gaussian copula and construct a three-dimensional $L_X-L_{UV}$ relation  by using the copula function. The comparisons between the three-dimensional and  standard $L_X-L_{UV}$ relations are made in Section 3. Finally, the conclusions are summarized  in Section 4.

~\\
\section{Three-dimensional $L_X-L_{UV}$ relation from copula}\label{Sec:Correlation&calibration}
\subsection{Copula}

Copula is proposed to describe the intercorrelation between statistical variables \citep{nelsen2007introduction}. It can join or ``couple'' multivariate distribution functions with one-dimensional marginal distribution functions.
Supposing that there are three variables $x_1$, $x_2$ and $x_3$ with the  marginal cumulative distribution functions (CDFs) being $F_1(x_1)$, $F_2(x_2)$ and $F_3(x_3)$, respectively,  the joint distribution function of these three variables can be described by using copula function $C$:
\begin{eqnarray}\label{H}
	H(x_1, x_2, x_3) = C(F_1(x_1), F_2(x_2), F_3(x_3))\, .
\end{eqnarray}
The key point of Eq.~(\ref{H}) is that by using copula one can model the dependence structure and the marginal distribution separately. In the following, we set $u_i=F_i(x_i)$ ($i=1,2,3$) for simplifying the expressions.  The joint probability density function (PDF) $h(x_1, x_2, x_3)$ of $H(x_1, x_2, x_3)$ can be obtained through
\begin{eqnarray}\label{h}
		h(x_1, x_2, x_3) &=&\frac{\partial^{3} H(x_1, x_2, x_3)}{\partial x_1 \partial x_2 \partial x_3} \nonumber \\
		&=&\frac{\partial^{3} C(u_1, u_2, u_3)}{\partial u_1 \partial u_2 \partial u_3} \frac{\partial u_1}{\partial x_1} \frac{\partial u_2}{\partial x_2} \frac{\partial u_3}{\partial x_3} \nonumber  \\
		&=&c(u_1, u_2, u_3) f_1(x_1) f_2(x_2) f_3(x_3) \, ,
\end{eqnarray}
where $f_i(x_i)$ ($i=1,2,3$) are the marginal PDFs of $F_i(x_i)$, and $c(u_1, u_2, u_3)$ is the density function of $C(F_1(x_1), F_2(x_2), F_3(x_3))$.

Two main copula families are elliptic copulas and Archimedean copulas.
The Gaussian copula belongs to a type of elliptic copulas and has a symmetric tail correlation. Since 
the Gaussian copula is the simplest and it can be analyzed analytically in many cases, we choose the Gaussian copula in our discussion and then we have
\begin{eqnarray}\label{Gaussian Copula}
	C(u_1, u_2, u_3 ; \bm{\theta})=\Psi_{3}\left[\Psi_{1}^{-1}(u_1), \Psi_{1}^{-1}(u_2), \Psi_{1}^{-1}(u_3) ; \bm{\theta} \right],
\end{eqnarray}
where $\bm{\theta}$ denotes the parameters of the copula function, and $\Psi_{3}$ and $\Psi_{1}$ are the standard three-dimensional and one-dimensional Gaussian CDFs, respectively,
which are defined as
\begin{eqnarray}
	\Psi_3 \left(\phi_{1}, \phi_{2}, \phi_{3} ; \bm{\theta} \right)
	&=& \int_{-\infty}^{\phi_{1}} \int_{-\infty}^{\phi_{2}} \int_{-\infty}^{\phi_{3}} \frac{1}{\sqrt{ (2\pi )^3 \mathrm{det}(\Sigma)}} \nonumber \\
	& \times &  \exp \left\{-\frac{1}{2}\left[ (\hat{\phi}_1, \hat{\phi}_2, \hat{\phi}_3)^T \Sigma^{-1} (\hat{\phi}_1, \hat{\phi}_2, \hat{\phi}_3)\right]\right \} \mathrm{d} \hat{\phi}_{1} \mathrm{d} \hat{\phi}_{2} \mathrm{d} \hat{\phi}_{3}\, ,
\end{eqnarray}
and
\begin{eqnarray}
	\Psi_1 \left(\phi \right) &=& \frac{1}{2} \mathrm{erfc} \left(-\frac{\phi}{\sqrt{2}} \right).
\end{eqnarray}
Here
\begin{eqnarray}\label{inverse function}
	\phi_i \equiv \Psi_{1}^{-1}(u_i)= -\sqrt{2} \,\mathrm{erfc}^{-1} \left(2u_i \right),
\end{eqnarray}
 $\Sigma$ is the covariance matrix, which relates  to the correlation coefficients $\bm{\theta}=\{\rho_{12}, \rho_{13}, \rho_{23}\}$ ($\Sigma_{ij}=\Sigma_{ji}=\rho_{ij},\Sigma_{ii}=1$), and
$\mathrm{erfc}$ and $\mathrm{erfc}^{-1}$ are complementary error function and its inverse, respectively.

The density function of the Gaussian copula can be calculated through
\begin{eqnarray}\label{Copula density}
	c(u_1, u_2, u_3 ; \bm{\theta})& =&\frac{\partial^{3} \Psi_{3}\left[\Psi_{1}^{-1}(u_1), \Psi_{1}^{-1}(u_2), \Psi_{1}^{-1}(u_3) ; \bm{\theta} \right]}{\partial u_1 \partial u_2 \partial u_3}  \nonumber  \\
	&=&\frac{1}{\sqrt{\mathrm{det} (\Sigma)}} \exp \left\{-\frac{1}{2}\left[\bm{\Psi}^{-1^{T}}\left(\Sigma^{-1}-\mathbf{I}\right) \bm{\Psi}^{-1}\right]\right\},
\end{eqnarray}
where $\bm{\Psi^{-1}}\equiv\{ \Psi_1^{-1}(u_1), \Psi_1^{-1}(u_2), \Psi_1^{-1}(u_3)\}$.
Then, the conditional PDF of $x_1$ denotes the probability of variable $x_1$ when $x_2$ and $x_3$ are fixed, which can be expressed as:
\begin{eqnarray}\label{f}
	f_{x_1}(x_1 | x_2, x_3 ; \bm{\theta})=\frac{c(u_1, u_2, u_3 ; \bm{\theta})f_1(x_1)f_2(x_2)f_3(x_3)}{c(u_2, u_3; \rho_{23})f_2(x_2)f_3(x_3)}=\frac{c(u_1, u_2, u_3 ; \bm{\theta})}{c(u_2, u_3 ; \rho_{23})}f_1(x_1)\, .
\end{eqnarray}
When the variable $x_i$ ($i=1, 2, 3$) obeys the Gaussian distribution with the mean value being $\mu_i$  and the standard deviation $\sigma_i$, its CDF $u_i$ can be expressed by
\begin{eqnarray}
	u_i =\frac{1}{2} \mathrm{erfc} \left(-\frac{x_i-\mu_i}{\sqrt{2}\sigma_i} \right) .
\end{eqnarray}
Thus, from the  Eq.~(\ref{inverse function}), one can obtain
\begin{eqnarray}\label{psi}
	\Psi_1^{-1} \left(u_i \right) =-\sqrt{2}~  \mathrm{erfc}^{-1} (2u_i)=-\sqrt{2} ~ \mathrm{erfc}^{-1} \left[2 \times \frac{1}{2} \mathrm{erfc} \left(-\frac{x_i-\mu_i}{\sqrt{2}\sigma_i} \right) \right] = \frac{x_i-\mu_i}{\sigma_i} \, .
\end{eqnarray}
Combining Eqs.~(\ref{Copula density}), (\ref{f}), and (\ref{psi}), we can obtain 
\begin{eqnarray}\label{fx}
	f_{x_1}(x_1 | x_2, x_3 ;  \boldsymbol{\theta})=\frac{1}{\sqrt{2 \pi }\sigma_{x_1 \mid x_2, x_3}} \exp \left[-\frac{1}{2} S(x_1, x_2, x_3; \boldsymbol{\theta})\right],
\end{eqnarray}
where $\sigma_{x_1 \mid x_2, x_3}$ is the standard deviation of $f_{x_1}$, and 
\begin{eqnarray}\label{S}
	 S(x_1, x_2, x_3; \boldsymbol{\theta})=\frac{\left[ \left(\rho _{23}^2-1\right) \hat{x}_1+\left(\rho _{12}-\rho _{13} \rho _{23}\right) \hat{x}_2+\left(\rho _{13}-\rho _{12} \rho _{23}\right) \hat{x}_3\right]^2}
{ \left(\rho _{23}^2-1\right) \left(\rho _{12}^2 +\rho _{13}^2+\rho _{23}^2-2 \rho _{13} \rho _{23} \rho _{12}-1\right) \sigma _1^2 \sigma _2^2 \sigma _3^2} 
\end{eqnarray}
with $\hat{x}_i\equiv (x_i-\mu_i)/\sigma_i$.  Since $x_1$ follows the Gaussian distribution, the maximum probability of $x_1$ can be obtained from $S(x_1, x_2, x_3 ; \boldsymbol{\theta})=0$. Thus, we can achieve the relation between variables $x_1$, $x_2$ and $x_3$  by solving $S(x_1, x_2, x_3 ; \boldsymbol{\theta})=0$, and find that the relation can be simplified to be
\begin{eqnarray}\label{relation}
	x_1=a+bx_2+cx_3\, ,
\end{eqnarray}
where
\begin{eqnarray}\label{coefficient}
	a&=&\frac{\mu _3 \left(\rho _{13}-\rho _{12} \rho _{23}\right) \sigma _1 \sigma _2+\mu _2 \left(\rho _{12}-\rho _{13} \rho _{23}\right) \sigma _1 \sigma _3}{\left(\rho _{23}^2-1\right) \sigma _2 \sigma _3}+\mu _1 \, , \nonumber \\
	b&=&\frac{\left(\rho _{13} \rho _{23}-\rho _{12}\right) \sigma _1}{\left(\rho _{23}^2-1\right) \sigma _2} \, , 
	\quad
	c=\frac{\left(\rho _{12} \rho _{23}-\rho _{13}\right) \sigma _1}{\left(\rho _{23}^2-1\right) \sigma _3} \, .
\end{eqnarray}

\subsection{Three-dimensional $L_X-L_{UV}$ relation}

\cite{Risaliti2015Hubble} have found that there is a nonlinear relation between the X-ray luminosity $L_X$ and UV luminosity $L_{UV}$  in observational data of quasars. The $L_X-L_{UV}$ relation has the form
\begin{eqnarray}\label{relation1}
	\log(L_X)=\beta + \gamma \log(L_{UV})\, ,
\end{eqnarray}
where $\beta$ and $\gamma$ are  two free parameters to be determined from the data, and 
$\log \equiv \log_{10}$.
Expressing the luminosity in terms of the flux, one can obtain
\begin{eqnarray}
	\log(F_X)=\beta '(z) + \gamma \log(F_{UV})\, ,
\end{eqnarray}
where $F_X$ and $F_{UV}$ are the X-ray and UV fluxes, respectively, and $\beta '(z)=2(\gamma-1)\log(d_L)+\beta + (\gamma-1)\log(4\pi) $. Here $d_L$ is the luminosity distance. We can fit parameters $\gamma$ and $\beta$ by using the quasar data, if   a concrete cosmological model is chosen.

\begin{figure}
	\centering	\includegraphics[width=0.7\textwidth]{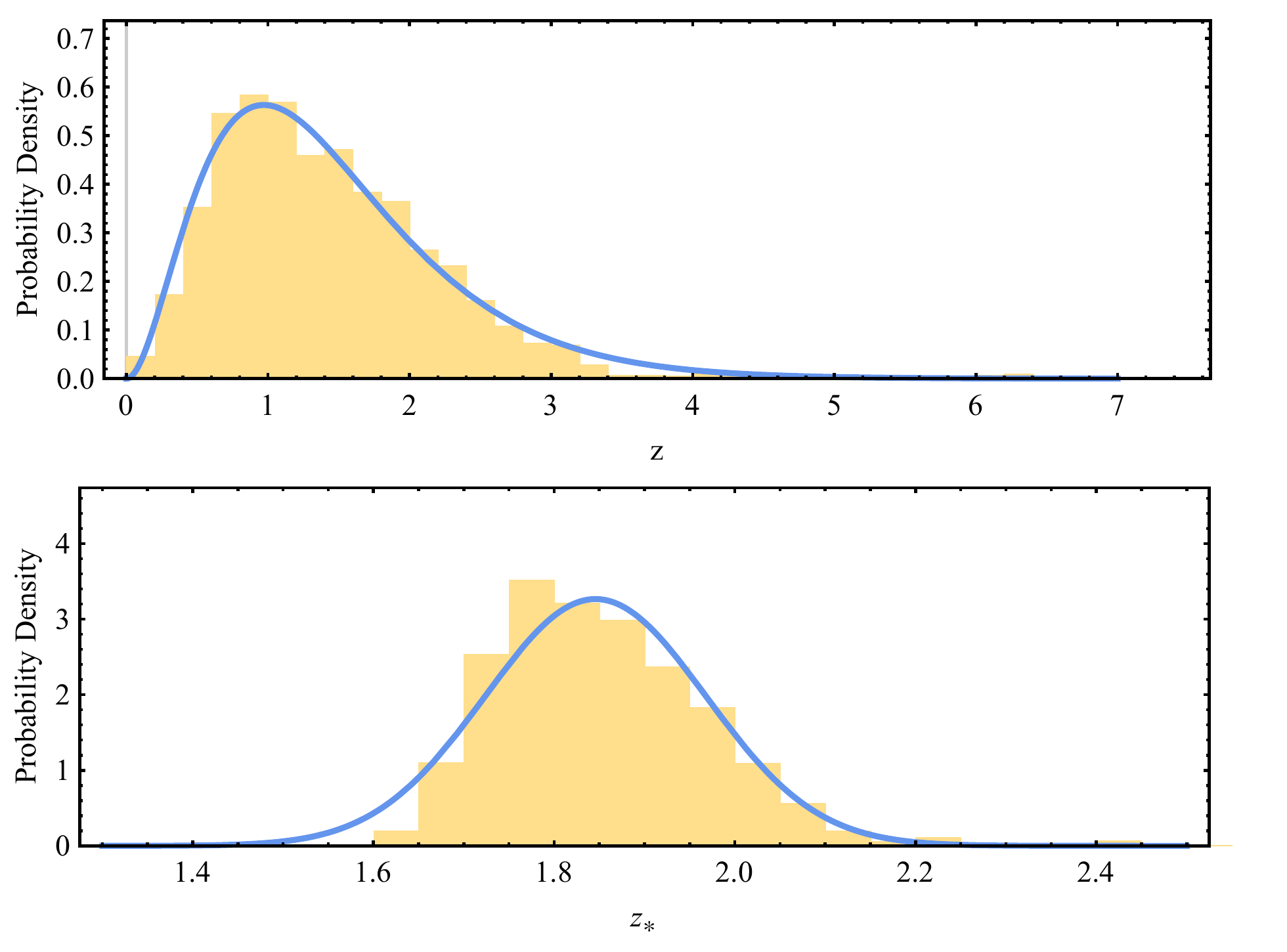}
	\caption{The histogram of   2421 quasar data points in the $z$ and $z_*$ spaces with the vertical axis representing  the probability density.  
	In the up and down panels, the solid lines represent the gamma distribution in the $z$ space and the Gaussian distribution in the $z_*$ space, respectively. 
		\label{fig1}
	}
\end{figure}

Now, we will use the copula function to construct the correlation between $L_X$, $L_{UV}$  and the redshift. First, we assume that both $\log(L_X)$ and $\log(L_{UV})$ follow  Gaussian distributions
with the distributional  functions being
\begin{eqnarray}\label{Eq9}
	f(x)=\frac{1}{\sqrt{2\pi}\sigma_x}e^{-\frac{(x-\bar{a}_x)^2}{2\sigma_x^2}}\,, \quad
	f(y)=\frac{1}{\sqrt{2\pi}\sigma_y}e^{-\frac{(y-\bar{a}_y)^2}{2\sigma_y^2}}\, .
\end{eqnarray}
 Here $x = \log(L_{UV})$, $y = \log(L_X)$,   $\bar{a}_x$ and  $\bar{a}_y$ represent the mean values, and $\sigma_x$ and $\sigma_y$ are the standard deviations. We have checked that these assumptions are very reasonable by performing a statistical distribution test (Cram\'{e}r-von Mises test \citep{Harald1928, Von1928}) on $x$ and $y$ based on the $\Lambda$CDM model.
Then, the redshift distribution of quasars  is considered.
In the upper panel of Fig.~\ref{fig1} we show  the probability density distribution of 2421 quasar data points~\citep{Lusso2020Quasars} in the redshift $z$ space, which  satisfies the gamma distribution apparently.
The probability density distribution is plotted  by requiring that  the total area enclosed by the curve and the $z$ axis is normalized to one. Thus, if  the value of the vertical axis is $P$ at redshift $z$,  the probability of the data in the redshift region $(z-\delta z, z+\delta z)$ is $2P\delta z$, where $\delta z$ is a very small variation and thus  the value of the vertical axis in the redshift region $(z-\delta z, z+\delta z)$ can be regarded as constant.
Since the log-transformation is a common way to transform the non-Gaussian distribution  into the Gaussian one, 
we consider the $z_*$ space, where  $z_*\equiv \ln(\bar{a}+z)$ with $\bar{a}$ being a constant and $\ln \equiv \log_e$,  and  find that the distribution of quasars can be described approximately by using a Gaussian distribution $f(z_*)$, which is shown in the down panel of Fig.~\ref{fig1}.  Thus,  $f(z_*)$ has the form
\begin{eqnarray}\label{redshift distribution}
	f(z_*)=\frac{1}{\sqrt{2\pi}\sigma_{z_{*}}}e^{-\frac{(z_*-\bar{a})^2}{2\sigma_{z_{*}}^2}}\,.
\end{eqnarray}
We obtain that $\bar{a}=5$ is a very good set after choosing some different values of $\bar{a}$ for comparison since it can lead  to very consistent fitting results in the following analysis. 

According to Eq.~(\ref{relation}), we obtain  the three-dimensional relation, which has the form  
\begin{eqnarray}\label{3D}
	\log(L_X)=\beta + \gamma \log(L_{UV})+\alpha \ln(\bar{a}+z)\,.
\end{eqnarray}
Apparently $\alpha$ is a new free parameter.  Eq.~(\ref{3D}) is different from the standard $L_X-L_{UV}$ relation (Eq.~(\ref{relation1})) with the addition of a redshift-dependent term. When $\alpha=0$ the standard relation is recovered.  When $\bar{a}=1$, our relation reduces to the one given in \citep{Dainotti2022} obtained by assuming that the luminosities of quasars are corrected by a redshift dependent function $(1+z)^\alpha$.  
Converting the luminosity to the  flux, one has
\begin{eqnarray}\label{relation2}
	\log(F_X)&=& \Phi(\log(F_{UV}),d_L) \nonumber \\
			&=& \beta' + \gamma \log(F_{UV}) +\alpha \ln(5+z)\, .
\end{eqnarray}
Eq.~(\ref{relation2}) is the main result of this Paper. In the following, we will discuss the viability of the three-dimensional $L_X-L_{UV}$ relation from copula.

~\\
\section{relation test}\label{Sec:Model test}

In this section, we use two different methods (low-redshift calibration and simultaneous fitting) to test the viability of the three-dimensional $L_X-L_{UV}$ relation given in Eq.~(\ref{relation2}) and make a comparison between the three-dimensional and standard $L_X-L_{UV}$ relations.

\subsection{Low-redshift calibration}
 Since $\beta'$ is dependent on the luminosity distance $d_L$, we choose the fiducial cosmological model  to give $d_L$. We consider  two models: the specially flat $\Lambda$CDM with ($H_0=70 ~\mathrm{km~s^{-1}~Mpc^{-1}}$, $\Omega_\mathrm{m0}=0.3$) and with ($H_0=70~ \mathrm{km~s^{-1}~Mpc^{-1}}$, $\Omega_\mathrm{m0}=0.4$), respectively.
Since the furthest type Ia supernovae reaches to redshift $z\sim2$ \citep{Scolnic2018}, we choose the quasar data within  the low-redshift region $z<2$   to determine the values of the coefficients in the three-dimensional and standard $L_X-L_{UV}$ relations. Extrapolating these results to the high-redshift ($z>2$) data, we can obtain the Hubble diagram of high-redshift quasar data at $z\sim7.5$. Finally, the high-redshift quasar data will be used to constrain the $\Lambda$CDM model.  If the value of $\Omega_\mathrm{m0}$  assumed in the fiducial model is compatible with the one from the high-redshift data, the low-redshift and high-redshift data give  consistent  results and thus we regard this relation as viable and then the quasars can be taken as  standard candles.

The data used in our analysis contain  2421 X-ray and UV flux measurements of quasars \citep{Lusso2020Quasars}, which cover the redshift range of $z \in  [0.009, 7.541]$. We use 1917 low-redshift data points for calibration.
The values of the coefficients in the  $L_X-L_{UV}$ relation  can be obtained by   minimizing $-\ln (\mathcal{L})$, where $\mathcal{L}$ is the D'Agostinis likelihood function \citep{D'Agostini2005Fits}
\begin{eqnarray}\label{Lc}
		&\mathcal{L}&(\delta,\beta,\gamma,\alpha)\propto\prod_{i} \frac{1}{\sqrt{2 \pi (\delta^2+\sigma_{i}^2 )}}
		\exp\left \{-\frac{[\log(F_X)_i-\Phi(\log(F_{UV}),d_L)_{i}]^2}{2\left(\delta^2+\sigma_{i}^2 \right) } \right\}\,.
\end{eqnarray}
Here $\sigma_{i}$ represent the measurement errors in $\log(F_X)$, $\delta$ is  an intrinsic dispersion,  function  $\Phi$ is defined in Eq.~(\ref{relation2}), and $d_L$ is the luminosity
 distance predicted by the $\Lambda$CDM model, which is calculated as
\begin{eqnarray}\label{DL}
		d_L=(1+z)\frac{1}{H_0}\int_{0}^{z}{\frac{dz'}{\sqrt{\Omega_\mathrm{m0}(1+z')^3+(1-\Omega_\mathrm{m0})}}}\, .
	\end{eqnarray}

In order to compare the three-dimensional  $L_X-L_{UV}$ relation and the standard one,   we use the Akaike information criterion (AIC) \citep{Akaike1974,Akaike1981} and the Bayes information criterion (BIC) \citep{Schwarz1978}, which are, respectively, defined as
\begin{eqnarray}\label{criterion}
\mathrm{AIC} & = & 2 p-2 \ln (\mathcal{L})\,, \\
\mathrm{BIC} & = & p \ln N-2 \ln (\mathcal{L})\, ,
\end{eqnarray}
where $p$ is the number of free parameters and $N$ is the number of data points. We need to calculate $\Delta$AIC(BIC) ($\Delta$AIC (BIC)$=$ AIC (BIC) $-$AIC$_\mathrm{min}$ (BIC$_\mathrm{min}$))  of the two relations when comparing them.  We have strong evidence against  the relation that has a large AIC(BIC) if $\Delta \mathrm{AIC(BIC)}>10$ \citep{Jeffreys1998}.

\begin{figure*}
	  \includegraphics[width=0.48\textwidth]
	  {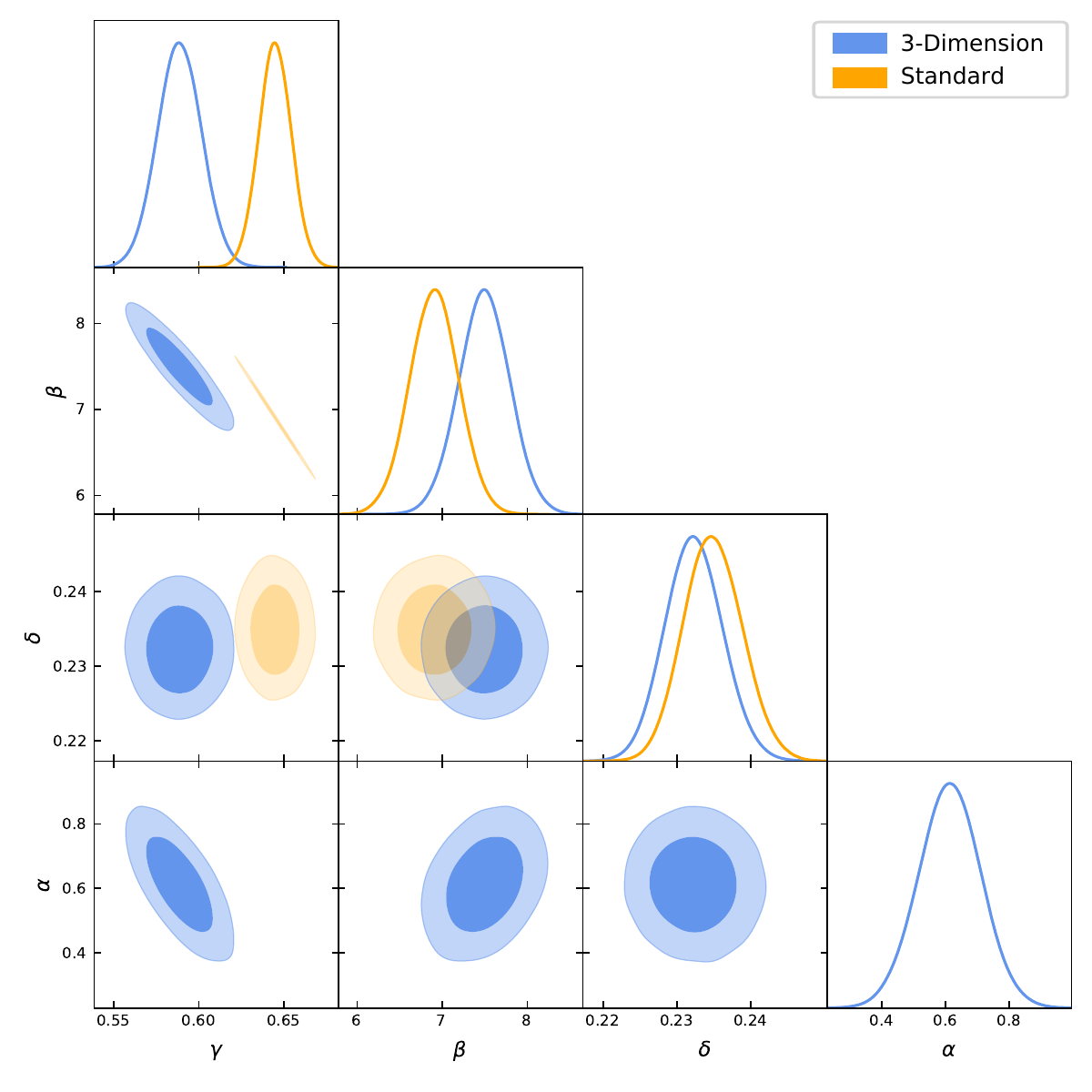}
	  \includegraphics[width=0.48\textwidth]
	  {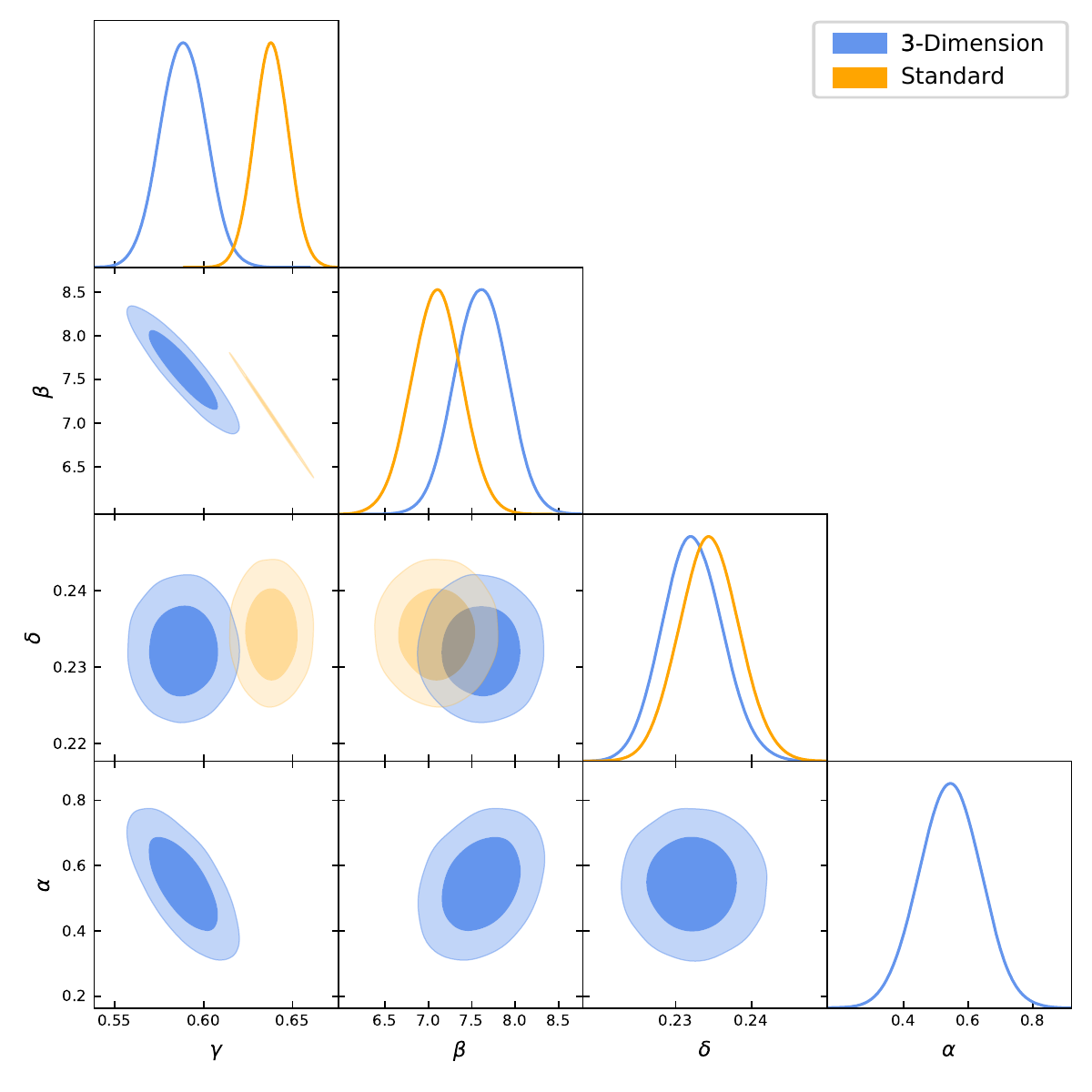}
	\caption{One-dimensional likelihood distributions and two-dimensional contours at 1$\sigma$ and 2$\sigma$ CLs.
	 $\alpha$, $\beta$, and $\gamma$ are the coefficients of the luminosity relation and $\delta$ is the intrinsic dispersion.
Left panel shows the result from the $\Lambda$CDM with $\Omega_{m0}= 0.3$, and  right panel shows the result  from the $\Lambda$CDM with $\Omega_{m0} = 0.4$. Blue and orange contours represent the three-dimensional and standard relations, respectively.
		\label{fig:LCDM-Low}
	}
\end{figure*}

The likelihood analysis is performed by using the Markov Chain Monte Carlo (MCMC) method as implemented in the $emcee$ package in $python$ 3.8 \citep{Foreman2013emcee}. Table~\ref{Tab2} and Figure~\ref{fig:LCDM-Low} show the results.
From them, we can conclude  that
\begin{itemize}

	\item The value of $\alpha$ deviates from  zero more than $3\sigma$,   which indicates that the observational data support apparently  the redshift-dependent luminosity relation.
	
	\item Since both  $\Delta$AIC  and $\Delta$BIC are larger than $20$, the three-dimensional relation from the copula function is favored strongly by the information criterions.

	\item The intrinsic dispersion  has a negligible difference for all cases. And the values of $\beta$  are almost independent of the cosmological models.
	
\end{itemize}

\begin{deluxetable*}{cccccc}
	\tablecaption{Marginalized one-dimensional best-fitting parameters with 1$\sigma$ confidence level from the low-redshift quasar data. \label{Tab2}}
	\tablewidth{0pt}
	\renewcommand{\arraystretch}{1.1}
	\tablehead{
		  &   \multicolumn{2}{c}{$\Omega_{m0}$ = 0.3}  & & \multicolumn{2}{c}{$\Omega_{m0}$ = 0.4 }  \\
		\cline{2-3} \cline{5-6}
		 & Standard & Three-Dimension & & Standard & Three-Dimension
	}
	\startdata
	$\delta$ & 0.235(0.004) & 0.232(0.004) & & 0.235(0.004) & 0.232(0.004)  \\
	$\beta$ & 6.906(0.292) & 7.502(0.302) & & 7.098(0.295) & 7.606(0.305)  \\
	$\gamma$ & 0.645(0.010) & 0.589(0.013) & & 0.638(0.010) & 0.589(0.013)  \\
	$\alpha$ & $-$ & 0.613(0.098) & & $-$ & 0.544(0.096)  \\
	\cline{1-6}
	$-2\ln\mathcal{L}$ &  $-$32.483 & $-$71.711 & & $-$39.661 & $-$71.346  \\
	$\Delta$AIC & 37.228 & $-$ & & 29.685 & $-$  \\
	$\Delta$BIC & 31.670 & $-$ & & 24.137 & $-$  \\
	\enddata
	\tablecomments{
		The values of $\sigma$ are in parentheses.}
\end{deluxetable*}

\begin{figure*}
	\center
	\includegraphics[width=0.495\textwidth]{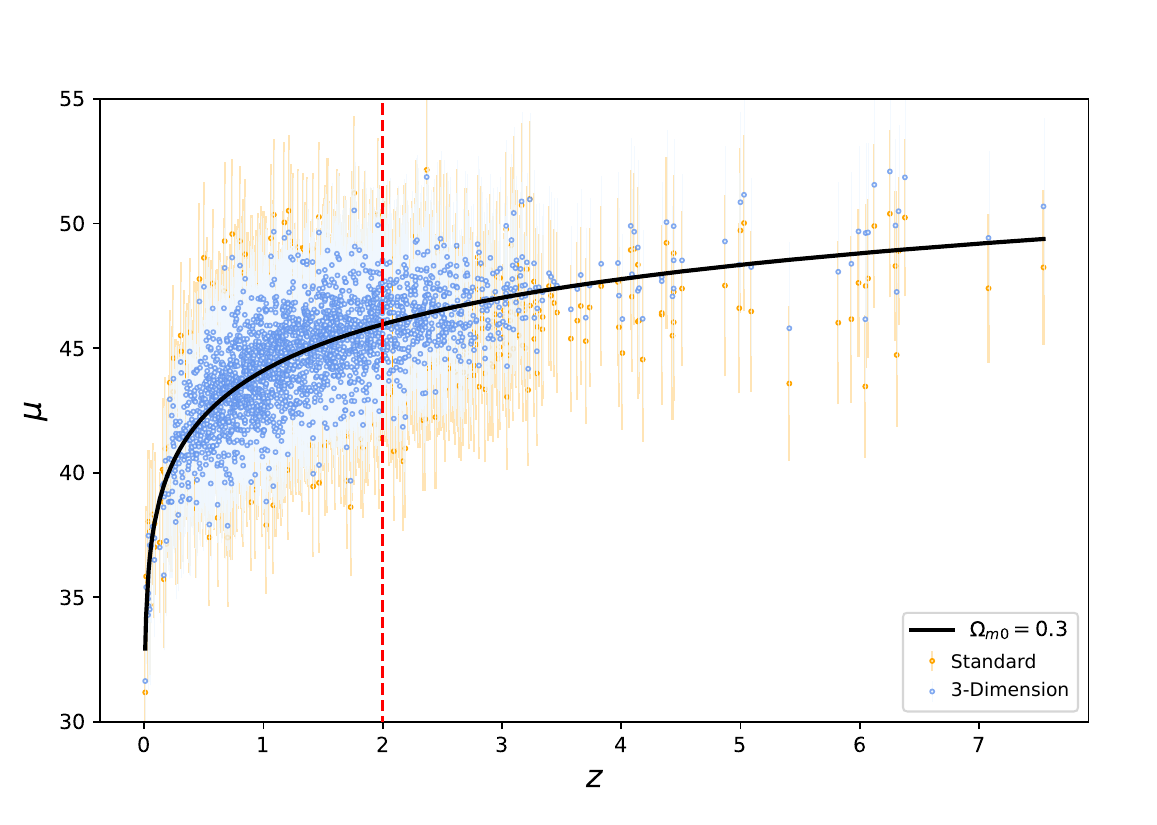}
	 \includegraphics[width=0.495\textwidth]{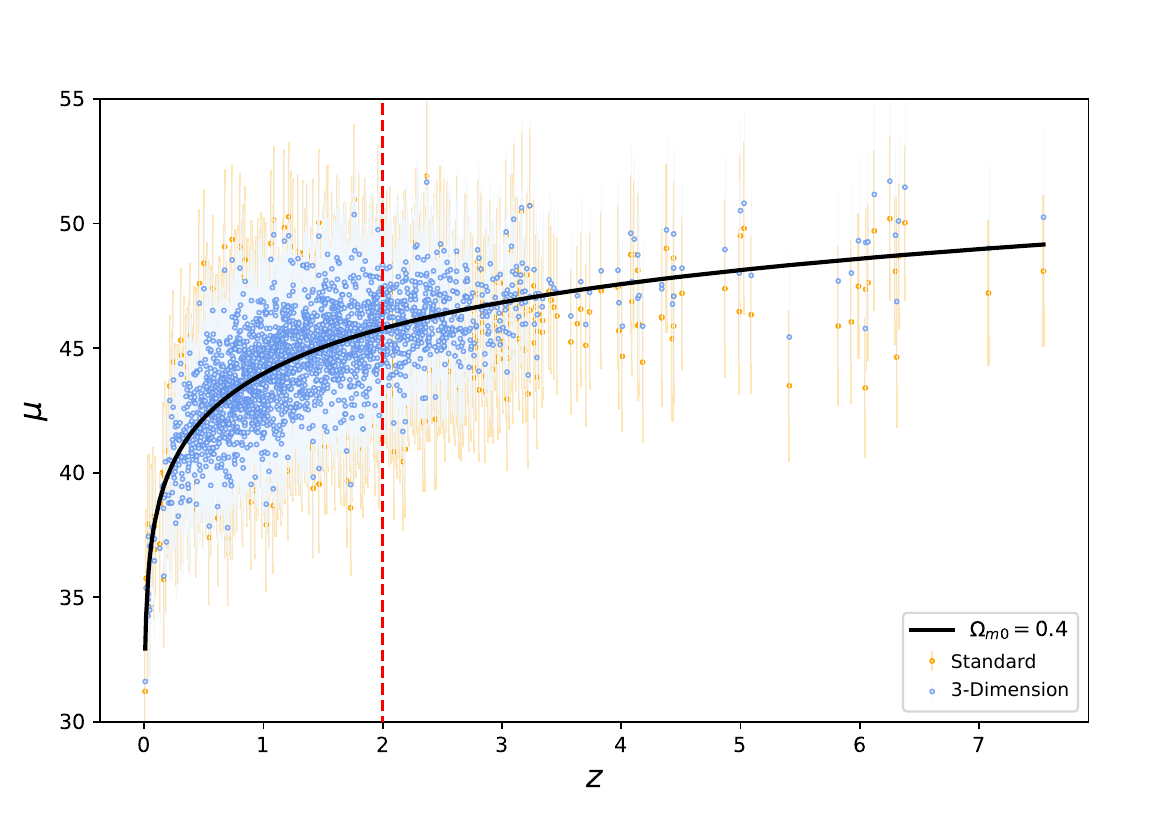}
	\caption{Hubble diagrams of quasars calibrated by low-redshift quasar data, which denotes the relation between the distance modulus $\mu$ and redshift $z$. The orange dots and the blue dots are derived from  the standard relation and the three-dimensional and redshift-evolutionary  one, respectively. The dashed red line corresponds to  $z=2$.
		\label{fig:Hubble}
	}
\end{figure*}

Extrapolating the values of the coefficients of the luminosity relations and the intrinsic dispersion from the low-redshift  data to the high-redshift samples, we can construct the Hubble diagram of quasars. The distance modulus and its errors are obtained from the formula given in the Appendix. In Fig.~\ref{fig:Hubble}, the Hubble diagrams of quasars obtained from two different relations (standard and three dimension) and two different fiducial  models are shown.  Apparently, the distance modulus from the three-dimensional relation is closer to the theoretical curve than that from the standard relation.

\begin{figure*}
	\center
	\includegraphics[width=0.49\textwidth]{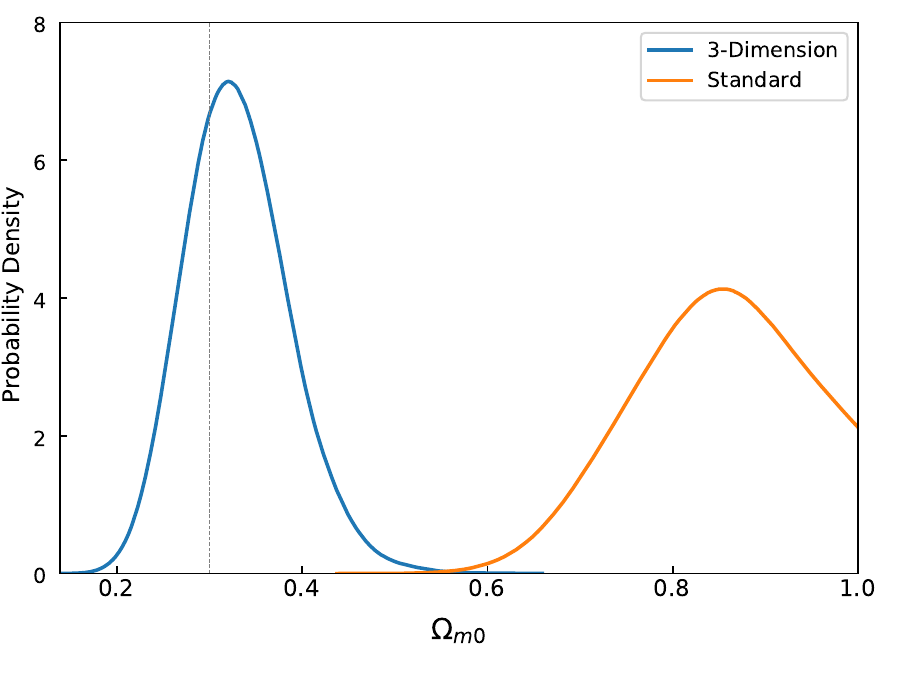}
	 \includegraphics[width=0.49\textwidth]{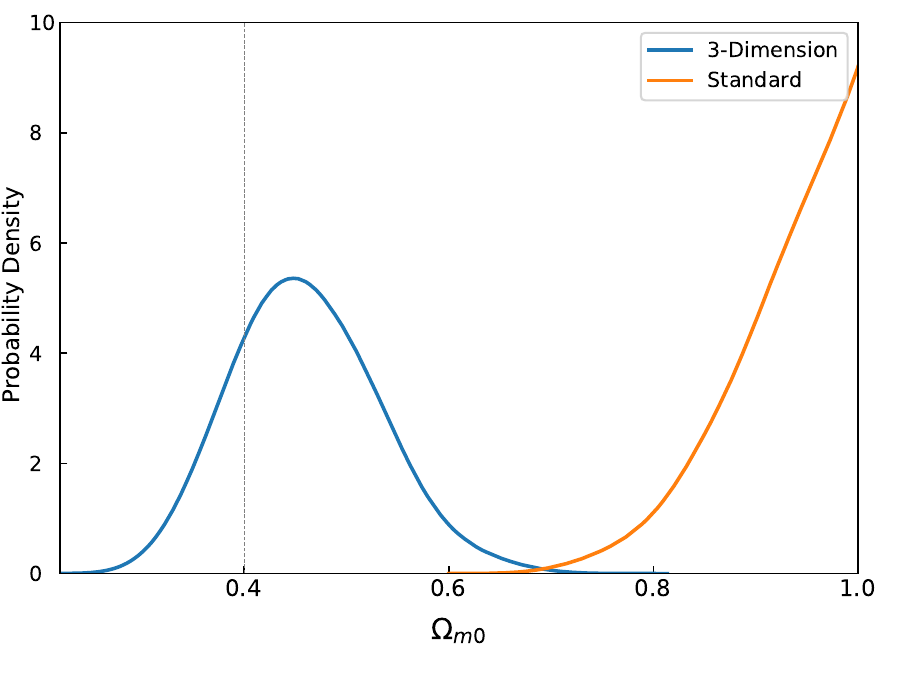}
	\caption{ Constraint on  $\Omega_{m0}$ of the  $\Lambda$CDM model from the high-redshift quasar data.
	The vertical axis represents the probability density.
	The  left and right panels represent the results from the $\Lambda$CDM model with $\Omega_\mathrm{m0}=0.3$ and $0.4$, respectively.
		\label{fig:LCDM-High}
	}
\end{figure*}

We use the distance modulus of high-redshift quasars to constrain the $\Lambda$CDM model  by minimizing $\chi ^2$
\begin{eqnarray}
\chi^{2} = \sum_{i = 1}^{504}\left[\frac{\mu_{o b s}\left(z_{i}\right)-\mu_{t h}\left(z_{i} \right)}{\sigma_{\mu_{i}}^{o b s}}\right]^{2}\, .
\end{eqnarray}
Here $\mu (z)=25+5\log \left[\frac{d_L(z)}{\mathrm{Mpc}}\right]$ is the distance modulus.
The results are shown in Fig.~\ref{fig:LCDM-High} and summarized  in Tab.~\ref{Tab3}. It is easy to see that the data from the three-dimensional relation can give a constraint on $\Omega_\mathrm{m0}$ consistent with that assumed in the fiducial model at the $1\sigma$ confidence level (CL), while the data from the standard relation cannot, and the deviations are larger than $3\sigma$. The AIC and BIC also favor strongly the three-dimensional relation. Therefore,
 we can conclude that   the three-dimensional luminosity relation from copula is obviously superior to the standard relation.   Using this three-dimensional relation, the quasars can be treated  as the standard candle  to probe the cosmic evolution.
 
\begin{figure*}
	\center
	\includegraphics[width=0.49\textwidth]{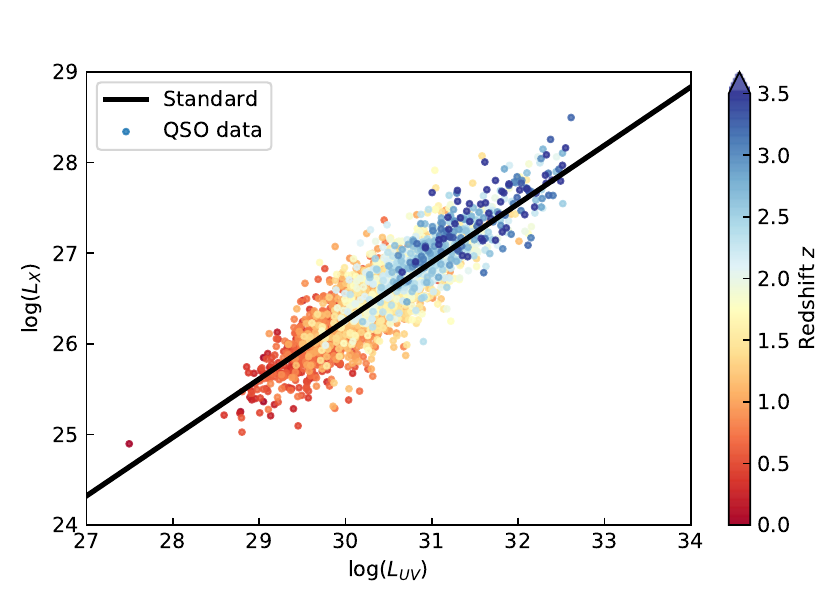}
	 \includegraphics[width=0.49\textwidth]{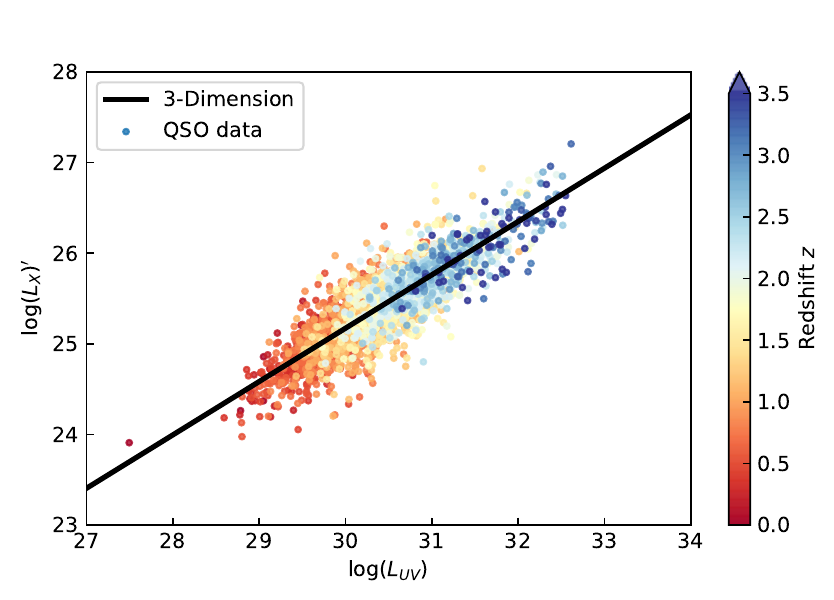}
	\caption{The comparison of  the relations of $\log(L_X)$-$\log(L_{UV})$ and $\log(L'_X)$-$\log(L_{UV})$ with quasar data points, where $\log(L_{X})'=\log(L_{X})-\alpha \ln(z+5)$. The solid lines are luminosity relations  calibrated by using the $\Lambda$CDM model with  $\Omega_{m0}=0.3$ from the  low-redshift quasar data. 
		\label{fig:luminosity relations}
	}
\end{figure*}

To compare two relations with data more clearly, we plot the quasar data points in the $\log(L_X)$-$\log(L_{UV})$ and $\log(L_X)'$-$\log(L_{UV})$ planes   in Fig.~\ref{fig:luminosity relations}, where $\log(L_{X})'\equiv \log(L_{X})-\alpha \ln(z+5)$.  The solid lines are luminosity relations calibrated by using the $\Lambda$CDM model with $\Omega_{m0}=0.3$ from the low-redshift data. It is easy to see that the three-dimensional and redshift-evolutionary relation is more consistent  with the high-redshift data than  the standard one.

\begin{deluxetable*}{cccccccc}
	\tablecaption{Marginalized one-dimensional best-fitting parameters with 1$\sigma$ CL from the  high redshift quasar data. \label{Tab3}}
	\tablewidth{0pt}
	\renewcommand{\arraystretch}{1.2}
	\tablehead{
		Data set & Calibration & Relation & $\Omega_{m0}$ & 68\%CL & $-2\ln\mathcal{L}$ & $\Delta$AIC & $\Delta$BIC
	}
	\startdata
	\multirow{4}{*}{High redshift} & \multirow{2}{*}{$\Omega_{m0}$ = 0.3} &Standard & 0.838 & ${}^{+0.110}_{-0.079}$ & 124.685 & 49.543 & 45.321 \\
	\cline{3-8}
	& &Three-Dimension & 0.326 & ${}^{+0.047}_{-0.063}$ & 73.142 & $-$ & $-$ \\
	\cline{2-8}
	& \multirow{2}{*}{$\Omega_{m0}$ = 0.4} &Standard & $>$0.902 & $-$ & 123.011 & 48.399 & 44.177 \\
	\cline{3-8}
	& &Three-Dimension & 0.459 & ${}^{+0.063}_{-0.079}$ & 72.612 & $-$ & $-$ \\
	\enddata
\end{deluxetable*}

\subsection{Simultaneous fitting }\label{Sec:Simultaneous Fitting}

In the previous subsection, the fiducial cosmological model is used to study the reliability of the two different  $L_X-L_{UV}$ relations. Here, we discard this condition. The coefficients of the relation and the cosmological parameters of $\Lambda$CDM model will be fitted simultaneously  by minimizing the value of the D'Agostinis likelihood function (Eq.~(\ref{Lc})).

\begin{figure}
	\center
	\includegraphics[width=0.8\textwidth]{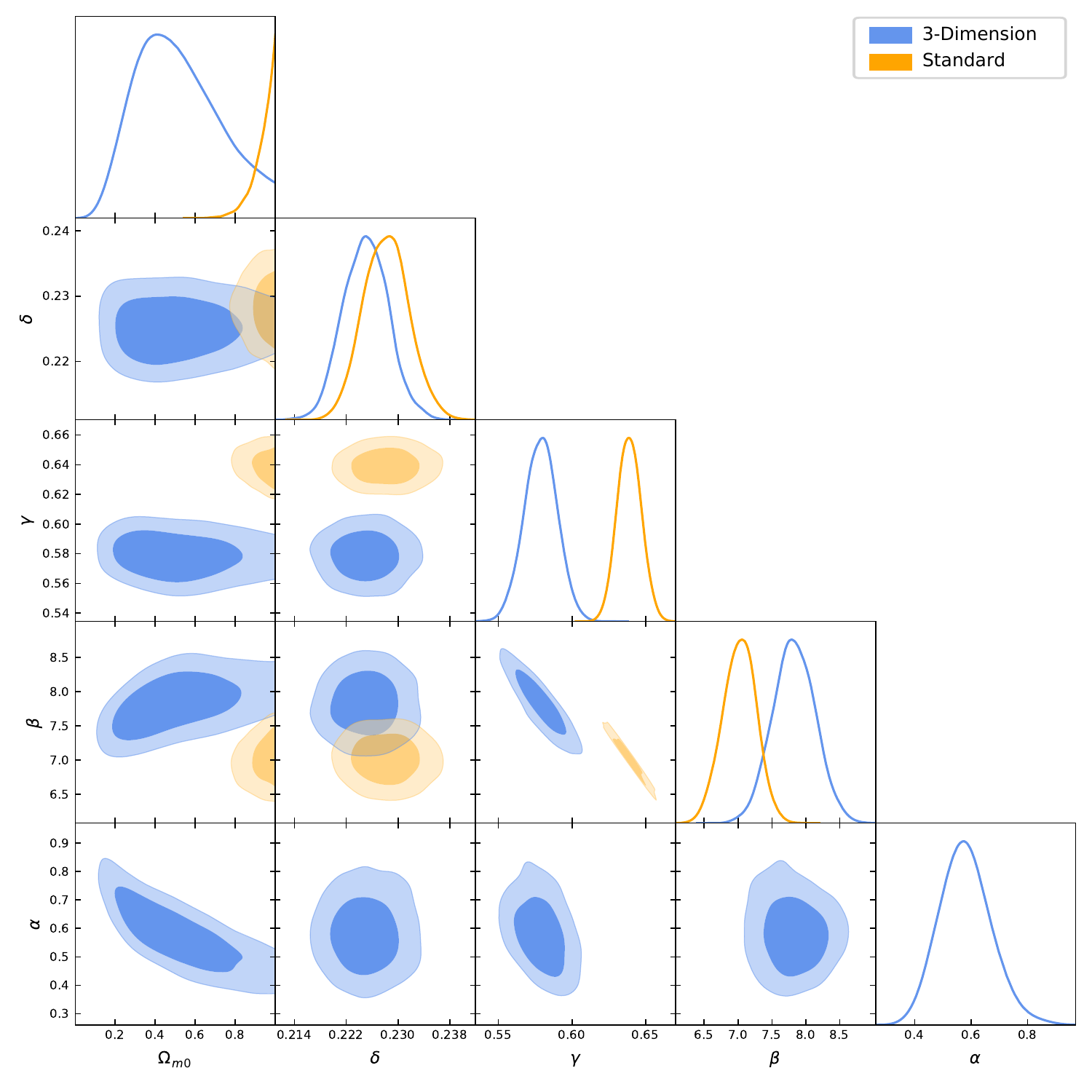}
	\caption{ One-dimensional likelihood distributions and two-dimensional contours at 1$\sigma$ and 2$\sigma$ CLs from quasar data with  the simultaneous fitting method. $\alpha$, $\beta$, and $\gamma$ are the coefficients of the luminosity relation, $\delta$ is the intrinsic dispersion and $\Omega_{m0}$ is the present dimensionless matter density parameter.
		\label{fig:LCDM-Global}
	}
\end{figure}

Figure~\ref{fig:LCDM-Global} shows one-dimensional probability density plots and contour plots of $\Omega_\mathrm{m0}$ and the coefficients of the two luminosity relations.  The marginalized mean values with $1\sigma$ CL are summarized in Tab.~\ref{Tab4}.
We find that the data favor apparently the redshift-evolutionary  relation since $\alpha$ deviates from zero more than $3\sigma$.   When the three-dimensional relation is used the quasar data can give an effective constraint on  $\Omega_\mathrm{m0}$ ($\Omega_\mathrm{m0}=0.510^{+0.163}_{-0.254})$, which is consistent with what were given by the current popular data including type Ia supernova and the cosmic microwave background radiation and so on. However the quasar data only gives a lower bound limit on $\Omega_\mathrm{m0}$  for the standard relation. Furthermore,  the AIC and BIC information criterions support strongly the three-dimensional  relation since both $\Delta$AIC and $\Delta$BIC are larger than 50.

\begin{deluxetable*}{c|ccccc|ccc}
	\tablecaption{Marginalized one-dimensional best-fitting parameters with 1$\sigma$ CL. \label{Tab4}}
	\renewcommand{\arraystretch}{1.1}
	\tablewidth{0pt}
	\tablehead{
		Relations & $\Omega_{m0}$ & $\delta$ & $\beta$ & $\gamma$ & $\alpha$ & $-2\ln\mathcal{L}$ & $\Delta$AIC & $\Delta$BIC
	}
	\startdata
	Standard & $>0.924$ & $0.228^{+0.004}_{-0.004}$ & $7.021^{+0.249}_{-0.249}$ & $0.639^{+0.008}_{-0.008}$ & $-$ & $-109.566$ & 65.100 & 59.308 \\
	Three-Dimension & $0.510^{+0.163}_{-0.254}$ & $0.225^{+0.004}_{-0.004}$ & $7.825^{+0.316}_{-0.316}$ & $0.579^{+0.011}_{-0.011}$ & $0.580^{+0.084}_{-0.099}$  & $-176.666$ & $-$ & $-$ \\
	\enddata
\end{deluxetable*}

~\\
\section{Conclusions}\label{Sec:Conclusion}

Using the powerful statistic tool called copula,  we construct a three-dimensional  $L_X-L_{UV}$ relation of quasars, which contains an extra redshift-dependent  term as opposed to the standard relation. We use two different methods to test the reliability of the relation from copula. One is to use the low-redshift quasar data to determine the values of relation coefficients after assuming the $\Lambda$CDM as the fiducial cosmological model, and extrapolate the results to the high-redshift data to give the Hubble diagram of quasars. Then these high-redshift data are used to constrain the $\Lambda$CDM model. By comparing the results from the high-redshift data with the ones given in the fiducial model, we can judge which relation is favored by quasar data. The other is to determine the coefficients of the luminosity relations and the cosmological parameters simultaneously from quasars. Two different methods give the same conclusion that the constraint on $\Omega_\mathrm{m0}$ from the three-dimensional  $L_X-L_{UV}$ relation is more consistent  with what was used in the fiducial cosmological model and obtained from other popular data than that from the standard relation. The observations favor a redshift-evolutionary  $L_X-L_{UV}$ relation more than $3\sigma$. According to the AIC and BIC information criterions, we find that the quasar data support strongly the  three-dimensional $L_X-L_{UV}$ relation. Our results indicate that  quasars can be regarded as the reliable indicator of the cosmic distance if the three-dimensional $L_X-L_{UV}$ relation is used to calibrate quasar data.

\acknowledgments
We appreciate very much the insightful comments and helpful suggestions by anonymous referees.
This work was supported in part by the NSFC under Grant Nos.~12275080, 12075084,  11690034,  11805063, and 12073069,   
 by the Science and Technology Innovation Plan of Hunan province under Grant No.~2017XK2019, and by the Guizhou Provincial  Science and Technology Foundation (QKHJC-ZK[2021] Key 020).

\appendix

\section{Distance modulus}\label{Appendix:Modulus}

The distance modulus $\mu(z)$ relates to the luminosity distance through \begin{eqnarray}\label{DMth}
		\mu (z)=25+5\log \left[\frac{d_L(z)}{\mathrm{Mpc}} \right ]\, .
\end{eqnarray}
Thus, if a cosmological model is assumed, the theoretical value of the distance modulus $\mu_\mathrm{t h}(z)$ can be obtained easily.
When the quasar data are considered,   the observed value of the distance modulus $\mu_\mathrm{obs} (z)$  can be deduced from Eq (\ref{relation2}) and Eq (\ref{DMth}) and has the form
\begin{eqnarray}\label{DMth2}
	\mu_\mathrm{obs} (z)=\frac{5}{2(\gamma-1)}\left[ \log(F_X)- \gamma \log(F_{UV})-\alpha \ln(5+z)- \beta ' \right]-97.447\, ,
\end{eqnarray}
where $\beta' = \beta+(\gamma-1)\log(4\pi)$.  Setting  $y=\log (F_X)$, $x=\log (F_{UV})$, one can  get the error of the distance modulus by using the  transfer formula
\begin{eqnarray}\label{sigma}
	\sigma_{\mu}^{2} & = &\delta^2+ \left(\frac{\partial \mu}{\partial \gamma}\right)^{2} \sigma_{\gamma}^{2}+\left(\frac{\partial \mu}{\partial \beta}\right)^{2} \sigma_{\beta}^{2}
	+\left(\frac{\partial \mu}{\partial \alpha}\right)^{2} \sigma_{\alpha}^{2}
	+\left(\frac{\partial \mu}{\partial y}\right)^{2}\sigma_{y}^{2}  \nonumber  \\
	&+& 2\sum_{i=1}^{4}\sum_{j=i+1}^{4}\left(\frac{\partial y(x;\bm{\bar{\theta}})}{\partial \bar{\theta}_i}\frac{\partial y(x;\bm{\bar{\theta}})}{\partial \bar{\theta}_j}\right) C_{ij}\,,
\end{eqnarray}
where $\bm{\bar{\theta}}=(\delta, \beta, \gamma, \alpha)$ and
\begin{eqnarray}\label{sigma2}
	\frac{\partial \mu}{\partial \gamma}&=&- \frac{\Gamma}{\gamma-1}\left [y-x-\alpha \ln(5+z)-\beta \right], \nonumber  \\
	\frac{\partial \mu}{\partial \beta}&=&-\Gamma, 	\qquad
	\frac{\partial \mu}{\partial y}=\Gamma, 	\qquad
	\frac{\partial \mu}{\partial \alpha}=-\Gamma \ln (5+z)\,.
\end{eqnarray}
Here $ \Gamma = \frac{5}{2(\gamma-1)}$. In Eq.~(\ref{sigma}),  the covariance matrix can be approximately evaluated from
\begin{eqnarray}
	\left(C^{-1}\right)_{ij}\left(\boldsymbol{ \bar{\theta}}\right) = \left.\frac{\partial^{2}\left[-\ln \mathcal{L}\left(\boldsymbol{\bar{\theta}}\right)\right]}{\partial \bar{\theta}_{i} \partial \bar{\theta}_{j}}\right|_{\boldsymbol{\bar{\theta}}  =  \boldsymbol{\bar{\theta}}_{0}}\,,
\end{eqnarray}
where $\boldsymbol{\bar{\theta}}_{0}$ denote the best-fitted values. Substituting Eq.~(\ref{sigma2}) into Eq.~(\ref{sigma}), one has   \begin{eqnarray}\label{errEiso}
	\sigma_{\mu}^2&=&
	\delta^2+\left [ \frac{\Gamma}{\gamma-1}\left (y-x-\alpha \ln(5+z)-\beta \right) \right]^2\sigma_\gamma^2
	+\Gamma^2 \sigma_\beta^2 +\Gamma^2 \sigma_y^2
	\nonumber \\
	&+& \left[\Gamma \ln (5+z) \right ]^2 \sigma_\alpha^2
	+2\sum_{i=1}^{4}\sum_{j=i+1}^{4}\left(\frac{\partial y(x;\bm{\bar{\theta}})}{\partial \bar{\theta}_i}\frac{\partial y(x;\bm{\bar{\theta}})}{\partial \bar{\theta}_j}\right) C_{ij}\, .
\end{eqnarray}



\begin{thebibliography}{}
\expandafter\ifx\csname natexlab\endcsname\relax\def\natexlab#1{#1}\fi
\providecommand{\url}[1]{\href{#1}{#1}}
\providecommand{\dodoi}[1]{doi:~\href{http://doi.org/#1}{\nolinkurl{#1}}}
\providecommand{\doeprint}[1]{\href{http://ascl.net/#1}{\nolinkurl{http://ascl.net/#1}}}
\providecommand{\doarXiv}[1]{\href{https://arxiv.org/abs/#1}{\nolinkurl{https://arxiv.org/abs/#1}}}


\bibitem[Adamcewicz \& Thrane(2022)]{Adamcewicz2022} Adamcewicz, C. \& Thrane, E.\ 2022,  \mnras, 517, 3928. \dodoi{10.1093/mnras/stac2961}

\bibitem[{Akaike(1974)}]{Akaike1974}
Akaike, H. 1974, ITAC, 19, 716

\bibitem[{Akaike(1981)}]{Akaike1981}
Akaike, H. 1981, J. Eon., 16, 3, \dodoi{10.1016/0304-4076(81)90071-3}

\bibitem[{{Ba{\~n}ados} {et~al.}(2018){Ba{\~n}ados}, {Venemans},
  {Mazzucchelli}, {Farina}, {Walter}, {Wang}, {Decarli}, {Stern}, {Fan},
  {Davies}, {Hennawi}, {Simcoe}, {Turner}, {Rix}, {Yang}, {Kelson}, {Rudie}, \&
  {Winters}}]{Ba2018million}
{Ba{\~n}ados}, E., {Venemans}, B.~P., {Mazzucchelli}, C., {et~al.} 2018, \nat,
  553, 473, \dodoi{10.1038/nature25180}
  
\bibitem[Banerjee et al.(2021)]{Banerjee2021} Banerjee, A., \'{O} Colg\'{a}in, E., Sasaki, M., Sheikh-Jabbari, M, M., \& Yang, T. 2021, Phys. Lett. B, 818, 136366, \dodoi{10.1016/j.physletb.2021.136366}

\bibitem[{{Baldwin}(1977)}]{Baldwin1977Luminosity}
{Baldwin}, J.~A. 1977, \apj, 214, 679, \dodoi{10.1086/155294}

\bibitem[{{Bargiacchi} {et~al.}(2021){Bargiacchi}, {Benetti}, {Capozziello},
  {Lusso}, {Risaliti}, \& {Signorini}}]{Bargiacchi2021Quasar}
{Bargiacchi}, G., {Benetti}, M., {Capozziello}, S., {et~al.} 2022, \mnras, 515, 1795. \dodoi{10.1093/mnras/stac1941}

\bibitem[{{Benabed} {et~al.}(2009){Benabed}, {Cardoso}, {Prunet}, \&
  {Hivon}}]{Benabed2009TEASING}
{Benabed}, K., {Cardoso}, J.~F., {Prunet}, S., \& {Hivon}, E. 2009, \mnras,
  400, 219, \dodoi{10.1111/j.1365-2966.2009.15202.x}

\bibitem[{{Cao} {et~al.}(2021){Cao}, {Ryan}, {Khadka}, \&
  {Ratra}}]{Cao2021Cosmological}
{Cao}, S., {Ryan}, J., {Khadka}, N., \& {Ratra}, B. 2021, \mnras, 501, 1520,
  \dodoi{10.1093/mnras/staa3748}

\bibitem[{{Cao} {et~al.}(2020){Cao}, {Ryan}, \& {Ratra}}]{Cao2020Cosmological}
{Cao}, S., {Ryan}, J., \& {Ratra}, B. 2020, \mnras, 497, 3191,
  \dodoi{10.1093/mnras/staa2190}

\bibitem[{{Cao} {et~al.}(2017){Cao}, {Zheng}, {Biesiada}, {Qi}, {Chen}, \&
  {Zhu}}]{Cao2017Ultra}
{Cao}, S., {Zheng}, X., {Biesiada}, M., {et~al.} 2017, \aap, 606, A15,
  \dodoi{10.1051/0004-6361/201730551}

\bibitem[{{Chen} \& {Ratra}(2003)}]{Chen2003Cosmological}
{Chen}, G., \& {Ratra}, B. 2003, \apj, 582, 586, \dodoi{10.1086/344786}


\bibitem[Cram\'{e}r(1928)]{Harald1928} Cram\'{e}r, H.\ 1928, Scandinavian Actuarial Journal, 13, 74. 

\bibitem[{{D'Agostini}(2005)}]{D'Agostini2005Fits}
{D'Agostini}, G. 2005, arXiv e-prints, physics/0511182.
\newblock \doarXiv{physics/0511182}
\bibitem[Dainotti et al.(2022)]{Dainotti2022} Dainotti, M.~G., Bargiacchi, G., Lenart, A. {\L}., et al.\ 2022, \apj, 931, 106. \dodoi{10.3847/1538-4357/ac6593}

\bibitem[{{Foreman-Mackey} {et~al.}(2013){Foreman-Mackey}, {Conley},
  {Meierjurgen Farr}, {Hogg}, {Lang}, {Marshall}, {Price-Whelan}, {Sanders}, \&
  {Zuntz}}]{Foreman2013emcee}
{Foreman-Mackey}, D., {Conley}, A., {Meierjurgen Farr}, W., {et~al.} 2013,
  {emcee: The MCMC Hammer}, Astrophysics Source Code Library, record
  ascl:1303.002.
\newblock \doeprint{1303.002}

\bibitem[Hu \& Wang(2022)]{Hu2022} Hu, J.~P. \& Wang, F.~Y.\ 2022, \aap, 661, A71. \dodoi{10.1051/0004-6361/202142162}

\bibitem[Hu \& Jeffreys(1998)]{Jeffreys1998} Jeffreys, H.\ 1998, The theory of probability (Oxford: Oxford Univ. Press)
\bibitem[{{Jiang} {et~al.}(2009){Jiang}, {Yeh}, {Chang}, \&
  {Hung}}]{Jiang2009Construction}
{Jiang}, I.-G., {Yeh}, L.-C., {Chang}, Y.-C., \& {Hung}, W.-L. 2009, \aj, 137,
  329, \dodoi{10.1088/0004-6256/137/1/329}

\bibitem[{{Khadka} \& {Ratra}(2020{\natexlab{a}})}]{Khadka2020Quasar}
{Khadka}, N., \& {Ratra}, B. 2020{\natexlab{a}}, \mnras, 492, 4456,
  \dodoi{10.1093/mnras/staa101}

\bibitem[{{Khadka} \& {Ratra}(2020{\natexlab{b}})}]{Khadka2020Using}
{Khadka}, N., \& {Ratra}, B. 2020{\natexlab{b}}, \mnras, 497, 263, \dodoi{10.1093/mnras/staa1855}

\bibitem[{{Khadka} \& {Ratra}(2021)}]{Khadka2021Determining}
{Khadka}, N., \& {Ratra}, B. 2021, \mnras, 502, 6140, \dodoi{10.1093/mnras/stab486}

\bibitem[{{Khadka} \& {Ratra}(2022)}]{Khadka2022quasar}
{Khadka}, N., \& {Ratra}, B. 2022, \mnras, 510, 2753, \dodoi{10.1093/mnras/stab3678}

\bibitem[{{Kilerci Eser} {et~al.}(2015){Kilerci Eser}, {Vestergaard},
  {Peterson}, {Denney}, \& {Bentz}}]{Kilerci2015Scatter}
{Kilerci Eser}, E., {Vestergaard}, M., {Peterson}, B.~M., {Denney}, K.~D., \&
  {Bentz}, M.~C. 2015, \apj, 801, 8, \dodoi{10.1088/0004-637X/801/1/8}

\bibitem[{{Koen}(2009)}]{Koen2009Confidence}
{Koen}, C. 2009, \mnras, 393, 1370, \dodoi{10.1111/j.1365-2966.2008.14116.x}

\bibitem[{{La Franca} {et~al.}(2014){La Franca}, {Bianchi}, {Ponti},
  {Branchini}, \& {Matt}}]{La2014New}
{La Franca}, F., {Bianchi}, S., {Ponti}, G., {Branchini}, E., \& {Matt}, G.
  2014, \apjl, 787, L12, \dodoi{10.1088/2041-8205/787/1/L12}

\bibitem[{{Li} {et~al.}(2021){Li}, {Keeley}, {Shafieloo}, {Zheng}, {Cao},
  {Biesiada}, \& {Zhu}}]{Li2021Hubble}
{Li}, X., {Keeley}, R.~E., {Shafieloo}, A., {et~al.} 2021, \mnras, 507, 919,
  \dodoi{10.1093/mnras/stab2154}
  
  \bibitem[Li {et~al.}(2022)]{Li2022}
  Li, Z.,  Huang, L., \&  Wang, J. 2022, \mnras, 517, 1901.
  \dodoi{10.1093/mnras/stac2735}
  

\bibitem[{{Lian} {et~al.}(2021){Lian}, {Cao}, {Biesiada}, {Chen}, {Zhang}, \&
  {Guo}}]{Lian2021Probing}
{Lian}, Y., {Cao}, S., {Biesiada}, M., {et~al.} 2021, \mnras, 505, 2111,
  \dodoi{10.1093/mnras/stab1373}

\bibitem[{{Liu} {et~al.}(2022{\natexlab{a}}){Liu}, {Chen}, {Liang}, {Yuan},
  {Yu}, \& {Wu}}]{Liu2022a}
{Liu}, Y., {Chen}, F., {Liang}, N., {et~al.} 2022{\natexlab{a}}, \apj, 931, 50,
  \dodoi{10.3847/1538-4357/ac66d3}

\bibitem[{{Liu} {et~al.}(2022{\natexlab{b}}){Liu}, {Liang}, {Xie}, {Yuan},
  {Yu}, \& {Wu}}]{Liu2022b}
{Liu}, Y., {Liang}, N., {Xie}, X., {et~al.} 2022{\natexlab{b}}, \apj, 935, 7,
  \dodoi{10.3847/1538-4357/ac7de5}

\bibitem[Lusso \& Risaliti(2016)]{Lusso2016} Lusso, E. \& Risaliti, G.\ 2016, \apj, 819, 154.   \dodoi{10.3847/0004-637X/819/2/154}
\bibitem[Lusso \& Risaliti(2017)]{Lusso2017} Lusso, E. \& Risaliti, G.\ 2017, \aap, 602, A79. \dodoi{10.1051/0004-6361/201630079}

\bibitem[Lusso et al.(2019)]{Lusso2019} Lusso, E., Piedipalumbo, E., Risaliti, G., et al.\ 2019, \aap, 628, L4. \dodoi{10.1051/0004-6361/201936223}

\bibitem[{{Lusso} {et~al.}(2020){Lusso}, {Risaliti}, {Nardini}, {Bargiacchi},
  {Benetti}, {Bisogni}, {Capozziello}, {Civano}, {Eggleston}, {Elvis},
  {Fabbiano}, {Gilli}, {Marconi}, {Paolillo}, {Piedipalumbo}, {Salvestrini},
  {Signorini}, \& {Vignali}}]{Lusso2020Quasars}
{Lusso}, E., {Risaliti}, G., {Nardini}, E., {et~al.} 2020, \aap, 642, A150,
  \dodoi{10.1051/0004-6361/202038899}

\bibitem[Lyke et al.(2020)]{Lyke2020} Lyke, B.~W., Higley, A.~N., McLane, J.~N., et al. 2020, \apjs, 250, 8, \dodoi{10.3847/1538-4365/aba623}

\bibitem[{{Melia}(2014)}]{Melia2014high}
{Melia}, F. 2014, \jcap, 2014, 027, \dodoi{10.1088/1475-7516/2014/01/027}

\bibitem[{{Mortlock} {et~al.}(2011){Mortlock}, {Warren}, {Venemans}, {Patel},
  {Hewett}, {McMahon}, {Simpson}, {Theuns}, {Gonz{\'a}les-Solares}, {Adamson},
  {Dye}, {Hambly}, {Hirst}, {Irwin}, {Kuiper}, {Lawrence}, \&
  {R{\"o}ttgering}}]{Mortlock2011luminous}
{Mortlock}, D.~J., {Warren}, S.~J., {Venemans}, B.~P., {et~al.} 2011, \nat,
  474, 616, \dodoi{10.1038/nature10159}

\bibitem[{Nelsen(2007)}]{nelsen2007introduction}
Nelsen, R.~B. 2007, An introduction to copulas (New York: Springer)

\bibitem[{{Osmer} \& {Shields}(1999)}]{Osmer1999Review}
{Osmer}, P.~S., \& {Shields}, J.~C. 1999, in Astronomical Society of the
  Pacific Conference Series, Vol. 162, Quasars and Cosmology, ed. G.~{Ferland}
  \& J.~{Baldwin}, 235

\bibitem[{{Paragi} {et~al.}(1999){Paragi}, {Frey}, {Gurvits}, {Kellermann},
  {Schilizzi}, {McMahon}, {Hook}, \& {Pauliny-Toth}}]{Paragi1999VLBI}
{Paragi}, Z., {Frey}, S., {Gurvits}, L.~I., {et~al.} 1999, \aap, 344, 51.
\newblock \doarXiv{astro-ph/9901396}

\bibitem[Petrosian et al.(2022)]{Petrosian2022} Petrosian, V., Singal, J., \& Mutchnick, S.\ 2022, \apjl, 935, L19. \dodoi{10.3847/2041-8213/ac85ac}

\bibitem[Qin et al.(2020)]{Qin2020} Qin, J., Yu, Y., \& Zhang, P.\ 2020, \apj, 897, 105. \dodoi{10.3847/1538-4357/ab952f}


\bibitem[{{Risaliti} \& {Lusso}(2015)}]{Risaliti2015Hubble}
{Risaliti}, G., \& {Lusso}, E. 2015, \apj, 815, 33,
  \dodoi{10.1088/0004-637X/815/1/33}

\bibitem[{{Risaliti} \& {Lusso}(2019)}]{Risaliti2019Cosmological}
{Risaliti}, G., \& {Lusso}, E. 2019, NatAs, 3, 272, \dodoi{10.1038/s41550-018-0657-z}

\bibitem[{{Ryan} {et~al.}(2019){Ryan}, {Chen}, \& {Ratra}}]{Ryan2019Baryon}
{Ryan}, J., {Chen}, Y., \& {Ratra}, B. 2019, \mnras, 488, 3844,
  \dodoi{10.1093/mnras/stz1966}

\bibitem[{{Sacchi} {et~al.}(2022){Sacchi}, {Risaliti}, {Signorini}, {Lusso},
  {Nardini}, {Bargiacchi}, {Bisogni}, {Civano}, {Elvis}, {Fabbiano}, {Gilli},
  {Trefoloni}, \& {Vignali}}]{Sacchi2022Quasars}
{Sacchi}, A., {Risaliti}, G., {Signorini}, M., {et~al.} 2022, \aap, 663, L7, \dodoi{10.1051/0004-6361/202243411}
\bibitem[Sato et al.(2010)]{Sato2010} Sato, M., Ichiki, K., \& Takeuchi, T.~T.\ 2010, \prl, 105, 251301. \dodoi{10.1103/PhysRevLett.105.251301}
\bibitem[Sato et al.(2011)]{Sato2011} Sato, M., Ichiki, K., \& Takeuchi, T.~T.\ 2011, \prd, 83, 023501. \dodoi{10.1103/PhysRevD.83.023501}

\bibitem[Scherrer et al.(2010)]{Scherrer2010} Scherrer, R.~J., Berlind, A.~A., Mao, Q., et al.\ 2010, \apjl, 708, L9. \dodoi{10.1088/2041-8205/708/1/L9}
\bibitem[{Schwarz(1978)}]{Schwarz1978}
Schwarz, G. 1978, AnSta, 6, 461 

\bibitem[Scolnic et al.(2018)]{Scolnic2018} Scolnic, D. M., Jones, D. O., Rest, A., et al. \href{https://doi.org/10.3847/1538-4357/aab9bb}{2018, \apj, 859, 101}
\bibitem[Singal et al.(2022)]{Singal2022} Singal, J., Mutchnick, S., \& Petrosian, V.\ 2022, \apj, 932, 111. \dodoi{10.3847/1538-4357/ac6f06}

\bibitem[Takeuchi(2010)]{Takeuchi2010} Takeuchi, T.~T.\ 2010, \mnras, 406, 1830. \dodoi{10.1111/j.1365-2966.2010.16778.x}
\bibitem[Takeuchi \& Kono(2020)]{Takeuchi2020} Takeuchi, T.~T. \& Kono, K.~T.\ 2020, \mnras, 498, 4365. \dodoi{10.1093/mnras/staa2558}
\bibitem[Takeuchi et al.(2013)]{Takeuchi2013} Takeuchi, T.~T., Sakurai, A., Yuan, F.-T., et al.\ 2013, EP\&S, 65, 281. \dodoi{10.5047/eps.2012.06.008}

\bibitem[Von Mises (1928)]{Von1928} Von Mises, R.\ 1928, Wahrscheinlichkeit, Statistik und Wahrheit (Wien: Springer)

\bibitem[{{Wang} {et~al.}(2021){Wang}, {Yang}, {Fan}, {Hennawi}, {Barth},
  {Banados}, {Bian}, {Boutsia}, {Connor}, {Davies}, {Decarli}, {Eilers},
  {Farina}, {Green}, {Jiang}, {Li}, {Mazzucchelli}, {Nanni}, {Schindler},
  {Venemans}, {Walter}, {Wu}, \& {Yue}}]{Wang2021Luminous}
{Wang}, F., {Yang}, J., {Fan}, X., {et~al.} 2021, \apjl, 907, L1,
  \dodoi{10.3847/2041-8213/abd8c6}

\bibitem[{{Wang} {et~al.}(2014){Wang}, {Du}, {Hu}, {Netzer}, {Bai}, {Lu},
  {Kaspi}, {Qiu}, {Li}, {Wang}, \& {SEAMBH
  Collaboration}}]{Wang2014Supermassive}
{Wang}, J.-M., {Du}, P., {Hu}, C., {et~al.} 2014, \apj, 793, 108,
  \dodoi{10.1088/0004-637X/793/2/108}

\bibitem[{{Watson} {et~al.}(2011){Watson}, {Denney}, {Vestergaard}, \&
  {Davis}}]{Watson2011New}
{Watson}, D., {Denney}, K.~D., {Vestergaard}, M., \& {Davis}, T.~M. 2011,
  \apjl, 740, L49, \dodoi{10.1088/2041-8205/740/2/L49}
\bibitem[Wei \& Melia(2020)]{Wei2020} Wei, J.-J. \& Melia, F.\ 2020, \apj, 888, 99, \dodoi{10.3847/1538-4357/ab5e7d}

\bibitem[Yang et al.(2020)]{Yang2020} Yang, T., Banerjee, A., \& \'{O} Colg\'{a}in, E. 2020, \prd, 102, 123532, \dodoi{10.1103/PhysRevD.102.123532}
  
\bibitem[{{Yuan} {et~al.}(2018){Yuan}, {Wang}, {Worrall}, {Zhang}, \&
  {Mao}}]{Yuan2018Determining}
{Yuan}, Z., {Wang}, J., {Worrall}, D.~M., {Zhang}, B.-B., \& {Mao}, J. 2018,
  \apjs, 239, 33, \dodoi{10.3847/1538-4365/aaed3b}



\end{thebibliography}

\end{document}